\documentstyle[preprint,aps]{revtex}
\begin{document}
%\draft
\title{On Neutrino Masses and Family Replication} 
\author{P. Q. Hung}
\address{Dept. of Physics, University of Virginia, Charlottesville,
Virginia 22901}
\date{}
\maketitle
\begin{abstract}
The old issue of why there are more than one family of quarks and leptons is
reinvestigated with an eye towards the use of anomaly as a tool
for constraining the number of families. It is found that, by assuming
the existence of right-handed neutrinos (which would imply that neutrinos
will have a mass) and a new chiral $SU(2)$ gauge theory,
strong constraints on the number of families can be obtained. In addition,
a model, based on that extra $SU(2)$, is constructed where
it is natural to have one ``very heavy'' fourth neutrino and
three almost degenerate light neutrinos whose masses are
all of the Dirac type.

\end{abstract}

%\narrowtext

The mystery of family replication can simply be paraphrased by 
the celebrated question of I. I. Rabi \cite{rabi} concerning the discovery of 
the muon: ``Who ordered that?''. It is fair to
say that, after 61 years, this question remains unanswered and is further
complicated by the discovery of a third family of quarks and leptons. 
It is reasonable to expect that any solution to this problem
will necessarily lie outside of the realm of the Standard Model (SM).

The second mystery has to do with the question of why neutrinos are massless or almost so. 
Any mass, in particular very tiny ones, will definitely point to physics beyond the SM. Most
recently, new results from the Super Kamiokande collaboration \cite{superk} appear 
to give credence to this possibility. Is there a symmetry which can explain the
smallness of neutrino masses, if present?

Is it possible somehow to envision a scenario in which these two mysteries are 
intimately intertwined? The answer might be yes.
If symmetries beyond the SM are gauge symmetries, one might be able to
exploit powerful constraints such as the freedom from both local (perturbative)
triangle and global (nonperturbative) anomalies.
To set the tone, we first briefly review some known facts about anomalies in the
SM.
 
Let us suppose that the SM were described by $SU(N)_c \otimes SU(2)_L \otimes
U(1)_Y$ instead of the usual $SU(3)_c \otimes SU(2)_L \otimes
U(1)_Y$. Quarks now have $N$ colors. If the hypercharge assignment
for the left-handed lepton doublet is $-1/2$, then the cancellation
of the triangle anomaly implies $Y_{q_L} = 1/2N$ for the left-handed quarks.
This alone says nothing about what $N$ should be. Some other argument
is needed.
%Each family of quarks and leptons would then transform as:
%$l_L = (1, 2, -1/2)$; $q_L = (N, 2, Y_{q_L})$; $e_R = (1, 1, -1)$;
%$u_R = (N, 1, Y_{u_R})$; $d_R = (N, 1, Y_{d_R})$. The charge generator
%is the usual $Q = T_3 + Y$. The cancellation of triangle anomaly in the SM
%is the well-known condition: $\sum_{i} Q_i = 0$ which implies
%that the hypercharges of the left-handed
%quarks are constrained to be $Y_{q_L} = 1/2N$. This alone says nothing
%about what the number of color should be. This is where the constraint from
%the global nonperturbative anomaly of chiral gauge $SU(2)$ comes in.

Witten \cite{witten} has made the observation that an $SU(2)$ gauge
theory with an odd number of doublets of Weyl fermions is 
anomalous in the sense that the
fermionic determinant $\sqrt{det\, i \not\!\nabla (A_{\mu})}$ changes sign under
a ``large'' gauge transformation $A_{\mu}^{U} = U^{-1} A_{\mu} U - iU^{-1}\partial_{\mu} U$.
This would make the partition function $Z$ vanish and the theory would be
ill-defined. This nonperturbative anomaly would then require an {\em even} number
of Weyl doublets in order for chiral $SU(2)$ to be consistent.
(This ambiguity in sign stems from the fact that
the fourth homotopy group $\Pi_4(SU(2)) = Z_2$.) Other groups that also have
similar non-trivial constraints are $Sp(N)$ for any $N$ and $O(N)$ for $N \leq 5$.

There are 1 + N chiral doublets in the SM (1 for the leptons
and N for the quarks). Witten's anomaly requires 1 + N to be {\em even}. In
consequence, the number of colors must be {\em odd}, i.e. N= 3, 5, ... (excluding 1).
Experimentally, we think that $N=3$, but more deeply it could come from
some Grand Unified scheme such as $SU(5)$.
It is amusing to envision a world in which N = 5 for example (quarks with charges
3/5 and -2/5). 

Let us apply this simple lesson to the family replication problem.
To do this, we shall make a number of assumptions. The first two are:

1) There is a family gauge group.

2) There are right-handed neutrinos.

Once the $SU(2)$ anomaly constraint of the SM has been satisfied,
there is nowhere else one can think of in using
it: one needs another chiral $SU(2)$. $SU(2)_R$ of the
well-known Left-Right model \cite{mohapatra1} comes to mind. 
However, it is easy to
see that both local and global anomalies are satisfied in the Left-Right
model for {\em each} family, just as it is in the SM. 

Let us now suppose that an extra $SU(2)$ does exist 
and that it is {\em not} the one associated with Left-Right models.
We now propose that it is
the right-handed neutrinos, and {\em only them}, that interact with this extra
$SU(2)$, to be denoted by $SU(2)_{\nu_R}$ from hereon, under which they 
transform as doublets
to be denoted by
$\eta_R =(\nu_R^{\alpha}, \tilde{\nu}_R^{\alpha})$,
where $\alpha$ refers to
possible family indices. At this point, it is irrelevant as to which member
of the doublet pairs with the left-handed neutrino to form a mass.
 
We shall assume the family group to be $SO(N_f)$ except for
$N_f = 6$ with a spinor representation. The reason for such a choice
is because, if $\eta_R =(\nu_R^{\alpha}, \tilde{\nu}_R^{\alpha})$ 
were to carry family indices,
the family group will not be vector-like (unlike QCD for example) and
one would encounter a problem with the usual perturbative triangle
anomaly unless the group is of the type $SO(n)$ ($n \neq 6$ if spinor
representations are used) and $E_6$. We are
not considering cases where the anomaly of different representations
cancel each other.
We shall assume that the usual left-handed and right-handed quarks and
leptons, as well as $\eta_R$, transform as a {\em vector}
representation of $SO(N_f)$, i.e. as a vector with $N_f$ components. This
is free from the triangle anomaly even if $N_f = 6$.

Our model is described by:
$SU(3)_c \otimes SU(2)_L \otimes U(1)_Y \otimes SO(N_f) \otimes SU(2)_{\nu_R};
q_L = (3, 2, 1/6, N_f, 1); u_R = (3, 1, 2/3, N_f, 1); d_R = (3, 1, -1/3, N_f, 1); 
l_L = (1, 2, -1/2, N_f, 1); e_R = (1, 1, -1, N_f, 1); \eta_R = (1, 1, 0, N_f, 2)$. 
This model does not have
any perturbative triangle anomaly . The only remaining
anomaly one should be concerned with is Witten's anomaly associated with
$SU(2)_{\nu_R}$. There are two possibilities:

1) With $\eta_R = (1, 1, 0, N_f, 2)$, global anomaly freedom
dictates $N_f$ = 2, 4, 6, 8, ...

2) If there exists a {\em family singlet} $\eta_R^{\prime} = (1, 1, 0, 1, 2)$
in addition to $\eta_R = (1, 1, 0, N_f, 2)$, global
anomaly freedom dictates $N_f$ = 3, 5, 7, ... (We exclude the one family case.)

These options will be referred to as the even and odd options respectively.
The arguments presented
below are not meant to ``fix'' the number of families but
are simply meant to indicate what the phenomenological implications
for each choice of $N_f$ might be. The correct choice is undoubtedly
fixed by a yet-unknown theory of everything. Here we simply attempt to see
what physical differences between the even and odd options might be
in the context of $SU(2)_{\nu_R}$, leaaving the deeper question to
someone else.

One might require that gauge couplings are free from
Landau singularities below the Planck scale in such a way that
unification of the SM gauge couplings, if it exists, occurs in the 
perturbative regime. This requirement rules out a large 
number of possible choices.
With this criterion, one can see that the even
option can only accomodate $N_f = 2, 4, 6$, while the odd option can only 
accomodate $N_f = 3, 5$. This is because for $N_f \geq 7$, one or more
gauge couplings will ``blow up'' before the Planck scale. There are
no reasons, in the absence of a deeper theory, to rule out any of 
the above choices. This will require other yet-unknown conditions.
The only thing one can say, in the context of our model, is that
%Without additional
%phenomenological constraints, one would not be able at this point to
%make the ``last cut''.
electroweak precision experiments appear to
rule out $N_f \geq 5$ and and that existential facts tell us that $N_f$ is at
least three. This leaves us with the choice $N_f=4$ for the even option
and $N_f = 3$ for the odd option.

There are real physical differences between the even and odd options. 
The former predicts the existence of a fourth generation whose consequences
have been recently discussed in Refs. \cite{hung1} and  \cite{hung2}. The
latter predicts the existence of a neutral family-singlet $\eta^{\prime}_R$
(doublet under $SU(2)_{\nu_R}$) which could probably have cosmological
consequences. In addition, as we point out below, it appears that the
even option prefers almost degenerate light neutrinos while the odd
option prefers a hierarchical structure for the light neutrinos.

Let us concentrate, in this paper, on neutrino masses within the even option
with $N_f =4$..
In addition to the fermions, there is the SM Higgs field which
transform as $\phi = (1, 2, 1/2, 1, 1)$. We shall require that {\em all}
fermions be endowed with a global
$B-L$ symmetry. Since we are dealing only with leptons in this
manuscript, a global $L$ symmetry is sufficient. This global $L$
symmetry would {\em prevent} a Majorana mass term of the type
$\eta^{i\,\alpha}_{R} \eta_{i\,\alpha\,R}$, where $i=1,2$ and
$\alpha = 1,..,4$. In fact, in this scenario, only Dirac masses
are allowed.
It the follows that the only 
Yukawa coupling that one can have 
(for the lepton sector) is
${\cal L}_Y = g_E \bar{l}_L^{\alpha} \phi e_{\alpha\, R} + h.c.$,
where $\alpha= 1,..,4$ is the family index. This is unsatisfactory
for two reasons. First, it gives equal masses to all four charged leptons.
Second, all four neutrinos are massless. We know that the charged
leptons are not degenerate in mass. Furthermore, the width of the Z boson
tells us that, if there were a fourth neutrino, its mass would have to be 
larger than half the Z mass. We therefore need to lift the degeneracy
among the charged leptons and to give a mass to the neutrinos
(at least to the fourth one). There are probably many ways to achieve this
and we shall present one of such scenarios here.

To achieve the above aim, we introduce the following
set of vector-like (heavy) fermions: $F_{L,R} = (1, 2, -1/2, 1, 1)$;
$M_{1 L,R} = (1, 1, -1, 1, 1)$ and $M_{2 L,R} = (1, 1, 0, 1, 1)$ as well as
the following scalar fields: $\Omega^{\alpha} = (1, 1, 0, 4, 1)$ and
$\rho_{i}^{\alpha} = (1, 1, 0, 4, 2)$, where $\alpha$ and $i$ are $SO(4)$
and $SU(2)_{\nu_R}$ indices respectively. Notice that $F_{L,R}$ and $M_{1L,R}$
are vector like under $SU(2)_L \otimes U(1)_Y$, while $M_{2L,R}$ is singlet
under everything. As a result, they can have {\em arbitrary} gauge-invariant
bare masses. 
 
The Yukawa part of the Lagrangian involving leptons can be
written as
\begin{eqnarray}
{\cal L}^Y_{Lepton}& =& g_E \bar{l}_L^{\alpha} \phi e_{\alpha\, R} +
G_1 \bar{l}^{\alpha}_{L} \Omega_{\alpha} F_{R} +
G_{M_1} \bar{F}_{L} \phi M_{1R}
G_{M_2} \bar{F}_{L} \tilde{\phi} M_{2R} +
G_2 \bar{M}_{1L} \Omega_{\alpha} e^{\alpha}_{R} + \nonumber \\
          &  &G_3 \bar{M}_{2L} \rho^{\alpha}_{i} \eta^{i}_{\alpha R}
+ M_F \bar{F}_L F_R + M_1 \bar{M}_{1L} M_{1R} +
M_2 \bar{M}_{2L} M_{2R} + h.c.
\end{eqnarray}
As we have stated above, the assumption of an unbroken $L$ symmetry
forbids the presence of Majorana mass terms. 
For reasons to be discussed below,
we shall assume that $M_{F,1,2} \sim$ the scale of the family
$SO(4)$ breaking. After integrating out the $F$, $M_1$, and $M_2$ fields,
the effective Lagrangian below $M_{F,1,2}$ reads
\begin{eqnarray}
{\cal L}^{Y,eff}_{Lepton}& =& g_E \bar{l}_L^{\alpha} \phi e_{\alpha \,R} +
G_E \bar{l}^{\alpha}_{L} (\Omega_{\alpha} \phi \Omega^{\beta}) e_{\beta \,R} +\nonumber \\
          &  &G_N \bar{l}^{\alpha}_{L}(\Omega_{\alpha} \tilde{\phi} \rho^{\beta}_{i}) 
\eta^{i}_{\beta \,R} + h.c. + \text{higher\, dimensional\, operators},
\end{eqnarray}
where
\begin{equation}
G_E =\frac{G_1 G_{M_1} G_2}{M_F M_1};\, G_N =\frac{G_1 G_{M_2} G_3}{M_F M_2}.
\end{equation} 
The above effective Lagrangian accomplishes two things: 1) If $\Omega$ develops a
non-vanishing vacuum expectation value (VEV), the second term in Eq. (2) might lift
the degeneracy of the charged lepton masses; 2) If both $\Omega$ and $\rho$ 
develop non-vanishing VEV's, the third term in Eq. (2) might give rise to a
Dirac neutrino mass. These extra mass terms are linked to the breakdown
of $SO(4) \otimes SU(2)_{\nu_R}$. 
We would like to now show that it is rather straightforward to obtain a
``heavy'' fourth neutrino and three very light ones. 

Let us assume: $<\Omega> = (0,0,0,V)$ and $<\rho> = (0,0,0,V^{\prime} \otimes s_1)$,
where $s_1 = \left( \begin{array}{c} 1 \\ 0 \end{array} \right)$. 
Notice that each component (under
$SO(4)$) of $\rho$ transforms as a doublet under $SU(2)_{\nu_R}$. If we denote
the 4th element of $\eta_R$ by $(N_R,\,\tilde{N}_{R})$, one can use
the above two VEV's along with $<\phi> = (0,\,v/\sqrt{2})$ in Eq.(2) to
write down a mass term for the 4th generation neutrino, namely
\begin{equation}
\tilde{G}_N \frac{v}{\sqrt{2}} \bar{N}_L N_R + h.c.;\,
\tilde{G}_N = G_1 G_{M_2} G_3 \frac{V\,V^{\prime}}{M_F\,M_2}.
\end{equation}
At {\em tree level}, all other neutrinos are massless. Two remarks are
in order concerning the 4th neutrino mass. First, it is a {\em Dirac} mass.
Second, the 4th neutrino could be rather {\em heavy}. In fact, it is not
unnatural to expect $G_1$, $G_{M_2}$ and $G_3$ to be of the order of unity.
In consequence, as long as
$ V\, V^{\prime}/M_F\,M_2$ = O(1) (either $V \sim V^{\prime} \sim M_F \sim
M_2$ or any other combination), one might expect the fourth neutrino to
be even as heavy as 175 GeV. Certainly, the LEP bound of $M_Z/2$ can easily
be satisfied. In our scenario where the mass is a {\em Dirac}
mass, it {\em is} natural to have a very heavy fourth neutrino and three
so-far-massless neutrinos. 
We show below that, although they are massless at tree level, they can
acquire a mass at one loop.

In the Higgs potential, there is a term which is crucial to the 
computation of the light neutrino masses, namely
\begin{equation}
{\cal L}_{\Omega \rho} = \lambda_{\Omega \rho} (\Omega^{\alpha} \rho_{\alpha\,i})
(\Omega^{\beta} \rho_{\beta}^{i}).
\end{equation}
It is beyond the scope of this paper to give a detailed discussion of the
pattern of symmetry breaking. Without loss of generality, we will assume 
that the physical scalars associated with $\Omega$ and $\rho$ have a mass 
$M_{\Omega}$ and $M_{\rho}$ respectively. The one-loop diagram which
contributes to all {\em four} neutrino {\em Dirac} masses is shown in Fig. 1,
where the coupling given in Eq. (1) has been used. To make the discussion
more transparent, we simplify the problem by assuming $M_F = M_2$. A
general case with $M_F \neq M_2$ can easily be dealt with.
It is easy to see from
Fig. 1 that this one-loop diagram gives a {\em common} mass to all
four neutrinos: $m_{\nu} = \tilde{G}_{\nu} \frac{v}{\sqrt{2}}$,
where
\begin{mathletters}
\begin{equation}
\tilde{G}_{\nu} =  G_1 G_{M_2} G_3 \frac{V\,V^{\prime}}{M^2} \frac{\lambda_{\Omega \rho}}
{16\,\pi^2}\,I(M,M_{\Omega}, M_{\rho}),
\end{equation}
\begin{eqnarray}
I(M,M_{\Omega}, M_{\rho})& = &M^2 \{\frac{2\,M^2}{(M_{\Omega}^2 + M^2)(M_{\rho}^2 - M^2)} +
\frac{M_{\rho}^2 (M_{\rho}^2 + M^2)}{M_{\Omega}^2 (M_{\rho}^2 - M^2)^2} \ln
(\frac{M_{\Omega}^2 + M_{\rho}^2}{M_{\rho}^2}) + \nonumber \\
& & \frac{M^2 (M^2 - 3\,M_{\rho}^2)}{M_{\Omega}^2 (M_{\rho}^2 - M^2)^2} \ln
(\frac{M_{\Omega}^2 + M^2}{M^2})\},
\end{eqnarray}
\end{mathletters}
and where we have already made use of the simplication $M_F = M_2 = M$.

At this stage, the mass of the first three neutrinos are simply $m_{\nu}
= \tilde{G}_{\nu}\, v/\sqrt{2}$ while that of the fourth generation neutrino
is $m_N = \tilde{G}_N\,v/\sqrt{2} + m_{\nu}$. To see that $m_{\nu}$ can
be much smaller than the first term in $m_N$ and hence much smaller than
$m_N$ itself, let us simply assume that $M_{\Omega} = M_{\rho} = M_S$
in Eq. (6b). The function $I(M, M_{\Omega}, M_{\rho})$ becomes $I(x=M^2/M_S^2)$.  
It is straightforward to see that, for finite values of $x$, the function
$I(x)$ has a {\em zero} at $x \approx 0.4965$. What this intriguing fact means 
is that $m_{\nu}$ can be {\em very small} when the heavy fermion mass $M$
approaches $\sqrt{0.4965}M_S$. In fact in that limit,
\begin{equation}
m_{\nu} / m_N = (\lambda_{\Omega \rho} /16 \pi^2) I(x), 
\end{equation}
where $x = M^2/M_S^2$.
Taking into
account the fact that $\lambda_{\Omega \rho} /16 \pi^2$ can be very small
itself, e.g. $\lambda_{\Omega \rho} /16 \pi^2 \approx 10^{-7}$ for
$\lambda_{\Omega \rho} \approx 2 \times 10^{-5}$, it is not hard to imagine
that $M$ can be close to but not necessarily equal to $\sqrt{0.4965}\, M_S$ for
$m_{\nu} / m_N$ to be much less than unity. Alternatively, one can have
$x \sim 1$, in which case $I(x) \sim -0.06$ and this will implies that
$\lambda_{\Omega \rho} \sim 10^{-7}$ in order to have light neutrino
masses in the eV region. If one wants, on the other hand, $\lambda_{\Omega \rho} \sim 1$,
this will then require that $x \sim 0.4965$.
With this simple discussion, one can see that
it is not unnatural in our scenario to have one ``very heavy'' neutrino
and three very light ones, without resorting to the famous see-saw
mechanism. One could have, for example, $m_N$ = O(100 GeV) and $m_{\nu}$ =
O(eV).

We now turn to the discussion- albeit a rather brief one- of neutrino oscillation
\cite{mohapatra2}.
In such a discussion, one of the relevant quantities is the mass difference:
$\Delta m_{ij}^2 = |m_{\nu_i}^2 - m_{\nu_j}^2|$, $i,j=1,2,3$, the other ones
being the oscillation angles. The oscillation angles are related to the 
leptonic ``CKM'' matrix defined by $V_{ij} = U_{l}^{\dagger} U_{\nu}$, 
where $U_{l}$ and $U_{\nu}$ are the matrices which diagonalize the 
charged and neutral lepton mass matrices respectively. 

To lift the degeneracy of the three light neutrinos, it seems obvious that
the remaining light family symmetry, $SO(3)$ (coming from $SO(4) \rightarrow
SO(3)$), has to be broken. To this end, let us assume:
$<\Omega> = (\tilde{v}_1,\tilde{v}_2,\tilde{v}_3,V)$ and 
$<\rho> = (\tilde{v}^{\prime}_1 \otimes s_1,\tilde{v}^{\prime}_2 \otimes s_1,
\tilde{v}^{\prime}_3 \otimes s_1,V^{\prime} \otimes s_1)$,
where $s_1 = \left( \begin{array}{c} 1 \\ 0 \end{array} \right)$.
It is beyond the scope of this paper to
study the full dynamics of the most general Higgs potential which should
constraint the values of the various VeVs. We have seen how a tree-level
mass for the 4th generation neutrino arised and how a common mass for 
all four neutrinos was obtained at one-loop level, when $\tilde{v}$'s
and $\tilde{v}^{\prime}$'s were assumed to be zero. The strategy now
is to ``crank up'' those VeVs from zero to some ``small'' values-
i.e. small compared with $V$ and $V^{\prime}$- and see what happens.
After substituting these two VEV's into Eq.(2),
one obtains a non-diagonal $4 \times 4$ neutrino mass matrix ${\cal M}$
whose elements are given by: ${\cal M}_{ii} = m_{\nu} + \tilde{G}_{ii} v/\sqrt{2}$,
${\cal M}_{44} = m_{\nu} + \tilde{G}_{N} v/\sqrt{2}$,
${\cal M}_{ij} = \tilde{G}_{ij} v/\sqrt{2}$,
${\cal M}_{ji} = \tilde{G}_{ji} v/\sqrt{2}$,
${\cal M}_{i4} = \tilde{G}_{i4} v/\sqrt{2}$,
${\cal M}_{4i} = \tilde{G}_{4i} v/\sqrt{2}$, with
$\tilde{G}_{ij} = G_1 G_{M_2} G_3 (\tilde{v}_{i} 
\tilde{v}^{\prime}_{j})/(M_F\,M_2)$,
$\tilde{G}_{i4} =  G_1 G_{M_2} G_3 (\tilde{v}_{i} V^{\prime})/(M_F\,M_2)$,
$\tilde{G}_{4i} =  G_1 G_{M_2} G_3 (V \tilde{v}^{\prime}_{i})/(M_F\,M_2)$,
where $m_{\nu}$ and $\tilde{G}_N$ are given by Eq. (6a) and Eq. (4) respectively
and $i,j = 1,2,3$. The analysis of such a mass matrix for arbitray
$\tilde{v}$ and $\tilde{v}^{\prime}$ is beyond
the scope of the paper. One can however still get a glimpse of various possibilities
by looking at  particular cases. The discussion which follows
is not meant to be complete nor realistic. A simple
example will be used to show how one can partially lift the degeneracy
and how one might proceed to construct a more realistic neutrino
mass matrix. 

As an example, let us put
$\tilde{v}_i = \tilde{v}$ and $\tilde{v}^{\prime}_i = \tilde{v}^{\prime}$, 
$\forall i$. We shall assume that the primary diagonal masses of
the light neutrinos come from the one-loop diagram (Eq.(7)). 
It is then interesting to notice that the mass splitting among the
three light neutrinos is related to the disparity in the breaking
scales $V^{(\prime)}$ (of $SO(4)$) and $\tilde{v}^{(\prime)}$ 
(of $SO(3)$), i.e. the
difference between the breaking of the full family symmetry and that
of the light family symmetry. Let us take a specific example to start
our discussion. Let us assume $m_N \sim 100$ GeV and $m_{\nu} \sim
1.4$ eV. (The reader is referred to our earlier discussion on the
reason why it is possible to have the previous masses.)
The elements (except for ${\cal M}_{44}$) of the neutrino 
mass matrix are expressed
in terms of two ratios: $r =\tilde{v}/V$ 
and $r^{\prime} = \tilde{v}^{\prime}/V^{\prime}$,
namely ${\cal M}_{i4}= m_N r$, ${\cal M}_{4i} = m_N 
r^{\prime}$, ${\cal M}_{ij} = {\cal M}_{ji}=
m_N r r^{\prime}$ while ${\cal M}_{ii}=m_{\nu} +
{\cal M}_{ij}$. A few remarks are in order. First, it is easy to check
that the {\em degeneracy} of the three light neutrinos is still
present if the two ratios, $r$ and $r^{\prime}$,
are equal, regardless of their magnitude. Second, if we wish the
bulk of the light neutrino mass to come from $m_{\nu}$ then
$r r^{\prime} \lesssim 10^{-12}$. As an example, we take
$r = 10^{-6}$ and $r^{\prime} = 10^{-7}$. (A similar answer
is found even for $r = 1$ and $r^{\prime} = 10^{-12}$.)
The diagonalization
of the neutrino mass matrix gives the following eigenvalues:
100 GeV, 1.400013439 eV, 1.4 eV, 1.4 eV. Notice that
$(1.400013439)^2 - 1.4^2 \sim 4 \times 10^{-5}$. This simple exercise simply
shows that one can lift the degeneracy of at least one of the three
light neutrinos in the ``right'' direction. To have a more ``realistic''
splitting, one has to examine the general case with $r_i$ and 
$r^{\prime}_i$, $i = 1,2,3$, all different from one another. This
will be presented elsewhere.
The lesson learned from the previous example is simple: the tiny mass
splitting among the three light neutrinos is related to the large disparity
in scales between the full family symmetry and that of the light families.

A preliminary investigation of the odd option, with three families
and one family singlet $\eta^{\prime}$, appears to indicate that
the preferred solution for the neutrino masses is that in which there
is a hierarchy $m_1 \ll m_2 \ll m_3$. This will be presented elsewhere.

Several issues which need to be investigated are: 1) The charged lepton 
sector whose diagonalization matrix will be an important component
of the leptonic ``CKM'' matrix; 2) A detailed study of neutrino oscillation
using (1) combined with the above analysis; 3)
the quark sector within the framework of the present model; 4) Additional roles
of the vector-like heavy fermions, $F, M_1, M_2$, other than simply
being ``the mothers of all neutrino masses''. In particular, it
would be interesting to study the quark counterparts of these fermions.

In summary, we have presented in this paper arguments showing how an
extra symmetry among right-handed neutrinos, $SU(2)_{\nu_R}$, might
shed light on the nature of family replication. We have also presented
a model, with four generations, in which it is not unnatural
to have one very massive fourth neutrinos and three very light ones.
Furthermore, all neutrino masses are {\em Dirac} masses, and
hence there should be no such phenomenon as neutrinoless double beta decay.

I would like to thank Andrzej Buras and Manfred Lindner
at the Technical University of Munich, Gino Isidori and the theory group
at LNF(Frascati), and Chris Hill and the theory group at Fermilab for
their warm hospitality where part of this work was carried out.
I would like also to thank Paul Frampton and Manfred Lindner for reading 
the manuscript and for useful comments.
This work is supported in parts by the US Department
of Energy under grant No. DE-A505-89ER40518.

\begin{figure}
\caption{Feynman graph showing the computation of $\tilde{G}_{\nu}$,
where $m_{\nu} = \tilde{G}_{\nu} \frac{v}{\sqrt{2}}$}
\end{figure}

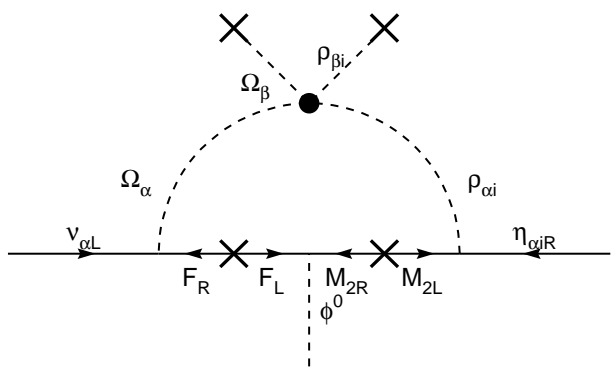
\end{document}